\documentclass[prb,twocolumn,superscriptaddress,balancelastpage]{revtex4}

\usepackage{graphicx,color}
\usepackage[colorlinks,plainpages=false,linkcolor=black,urlcolor=blue,citecolor=blue,pdfpagemode=UseNone,pdfstartview=FitH]{hyperref}

\begin{document}

\title{Photoemission induced gating of topological insulator}

\author{A.~A.~Kordyuk}
\affiliation{Institute for Solid State Research, IFW-Dresden, P.O.Box 270116, D-01171 Dresden, Germany}
\affiliation{Institute of Metal Physics of National Academy of Sciences of Ukraine, 03142 Kyiv,  Ukraine}

\author{T.~K.~Kim}
\author{V.~B.~Zabolotnyy}
\author{D.~V.~Evtushinsky}
\author{M.~Bauch}
\author{C.~Hess}
\author{B.~B\"{u}chner}
\affiliation{Institute for Solid State Research, IFW-Dresden, P.O.Box 270116, D-01171 Dresden, Germany}

\author{H.~Berger}
\affiliation{Institute of Physics of Complex Matter, EPFL, 1015 Lausanne, Switzerland}

\author{S.~V.~Borisenko}
\affiliation{Institute for Solid State Research, IFW-Dresden, P.O.Box 270116, D-01171 Dresden, Germany}

\begin{abstract}
The recently discovered topological insulators exhibit topologically protected metallic surface states which are interesting from the fundamental point of view and could be useful for various applications if an appropriate electronic gating can be realized. Our photoemission study of Cu intercalated Bi$_2$Se$_3$ shows that the surface states occupancy in this material can be tuned by changing the photon energy and understood as a photoemission induced gating effect. Our finding provides an effective tool to investigate the new physics coming from the topological surface states and suggests the intercalation as a recipe for synthesis of the material suitable for electronic applications.
\end{abstract}

%\preprint{\textit{xxx}}
\maketitle

The cone-like dispersion of the surface states with the spin degenerate Dirac point, recently observed in the photoemission experiments \cite{Xia,Hsieh,Chen}, is now considered as a hallmark of the topological insulators \cite{Moore,Kane,Fu}. In this sense, the compound Bi$_2$Se$_3$ with a single Dirac cone in the Brillouin zone \cite{Xia,Hsieh} can be considered as the most elementary one in the rapidly growing family of topological insulators \cite{Moore}. This, together with easy tunable charge carrier concentration by the surface doping \cite{Hsieh,Chen}, and realization of the superconductivity with Cu intercalation \cite{Hor,Wray}, make the Bi$_2$Se$_3$ class of materials the most promising for applications in spintronic and computing technologies \cite{Xia,Hsieh,Chen,Wilczek,Nayak,Zutic}. The main obstacle here is that unlike graphene \cite{Geim} the 3D topological insulators cannot be very easily tuned to the zero carrier density regime through standard electrical gating \cite{Xia,Hsieh}, though some successful attempts have been already reported \cite{Steinberg}. In this letter, using the ultra low temperature synchrotron based angle resolved photoemission spectroscopy (ARPES) in a wide photon energy range from $hv$ = 20 to 110 eV, we report the observation of the photoemission induced gating of Cu intercalated Bi$_2$Se$_3$. This finding opens a possibility to explore directly and on the same sample the effect of the surface state occupancy on the low energy electronic structure, as well as suggests the intercalation as a recipe for synthesis of the topological insulators suitable for electronic applications.

\begin{figure}[b]
\begin{center}
\includegraphics[width=8cm]{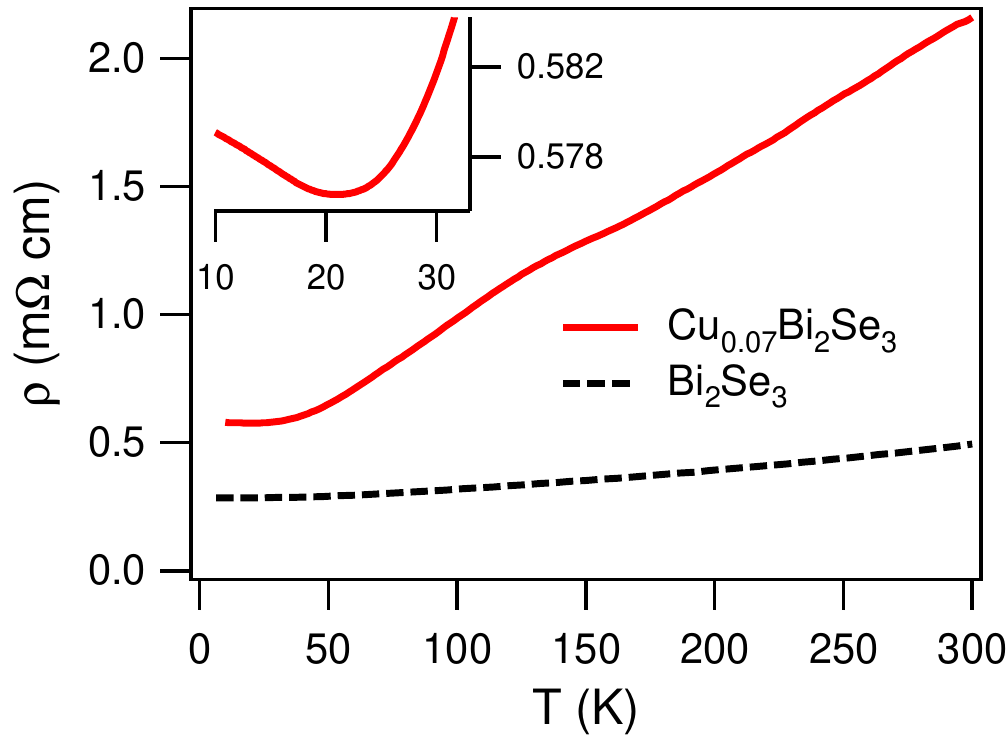}
\caption{
In-plane resistivity of the pure (black dashed line) and Cu intercalated (red solid line) Bi$_2$Se$_3$ crystals; the low temperature region of the later is zoomed in the inset.
\label{fig0}}
\end{center}
\end{figure}

\begin{figure*}
\begin{center}
\includegraphics[width=1\textwidth]{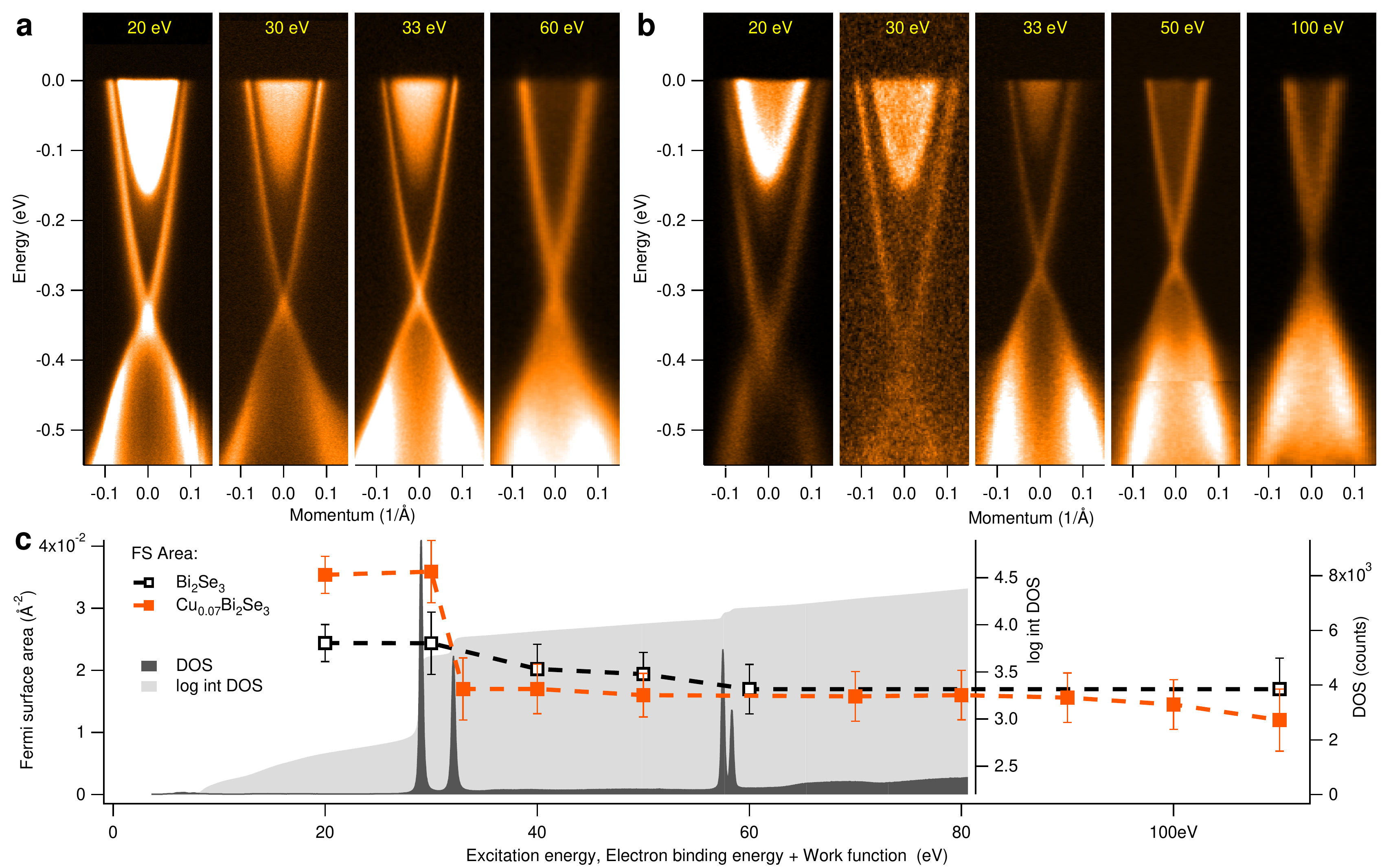}
\caption{
Excitation energy dependence of the surface states occupancy. A weak decrease of the binding energy of the Dirac point in Bi$_2$Se$_3$ (a) is contrasted to a 150 meV step-like change in Cu$_{0.07}$Bi$_{2}$Se$_{3}$ (b). (c) The step at 30 eV in the surface states occupancy, shown in units of the Fermi surface area, corresponds to the lowest photon energy enabling photoemission from Bi 5d core levels and consequent manifold increase of the photocurrent (see underlying plot of integral density of states, DOS).
\label{fig1}}
\end{center}
\end{figure*}

Single crystals of Bi$_2$Se$_3$ and Cu$_{0.07}$Bi$_{2}$Se$_3$ were grown using the Bridgman method starting with high-purity Bi, Cu, and Se elements. Samples in the form of rectangles with typical dimensions of 5$\,\times\,$5$\,\times\,$1 mm$^3$ were cut for measurements. According to EDX analysis, the Bi$_2$Se$_3$ samples from different batches show different amount of Se in the range ±1\%. The sample we measured by ARPES is characterized by typical \cite{Hyde} metallic in-plane resistivity with $\rho_{10K}$ = 0.3 m$\Omega\,$cm and $\rho_{300K}$ = 0.5 m$\Omega\,$cm. The Cu-doped crystals, that exhibit the large photo-induced gating effect, have been characterized by EDX as Cu$_{0.07}$Bi$_{2}$Se$_3$. They reveal much higher resistivity, as shown in Fig.~1, with semiconductor-like upturn at low temperature (see the inset). The out-of-plane resistivity for the Cu-doped samples reveals purely semiconducting behavior, increasing monotonically with lowering temperature from 10--18~m$\Omega\,$cm at 300~K up to 7--38~$\Omega\,$cm at 10~K. It is known \cite{Vasko} that Cu doping of Bi$_2$Se$_3$ can be realized in two ways: (1) singly ionized interstitial Cu atoms, which go between Se or Bi layers \cite{Vasko}, act as donors; (2) its substitutional defects on bismuth sites, which carry double negative charge, act as acceptors. Both the composition and resistivity data indicate that Cu, in the studied samples, appears mainly as interstitial atoms. Also, we observe the increase of the electronic occupation of the surface states in Cu$_{0.07}$Bi$_{2}$Se$_3$ comparing to Bi$_2$Se$_3$, that is consistent with the Cu intercalation.

ARPES experiments have been performed at the ``1$^3$" beamline at BESSY equipped with SES 4000 analyser and 3He cryo-manipulator with the base temperature on the sample less than 1 K \cite{Borisenko}. The data are sumarized in Fig.~2. The presented spectra were recorded along the cuts through the centre of the Brillouin zone close to $\Gamma$M direction. The band position was reproducible with 20 meV accuracy when measured at 1 K and under 8$\times10^{-11}$ mBar pressure on the time scale of the experiment about 6 hours as well as after keeping the cleaved sample at room temperature and at 2$\times10^{-10}$ mBar during 34 hours. This reproducibility can be seen even from the presented data. For example, the protocol of recording the spectra for Cu$_{0.07}$Bi$_{2}$Se$_{3}$ presented in Fig.~2b is the following: sample cleavege; +2h 40min; 50 eV spectrum; +40 min; 33 eV spectrum; +2h 40min; 20 eV spectrum; +1h; 50 eV spectrum again; +1h; 100 eV spectrum; +34h; 30 eV spectrum. One should note that for several Bi$_2$Se$_3$ samples measured at higher temperatures (10-30 K) we indeed observe a change in the binding energy of the Dirac point as a function of exposition time, about 30 meV during 20 min, that is in agreement with other reports \cite{Hsieh}. However, the spectra are recovered when the position of the light spot on the sample surface is changed.

\begin{figure*}
\begin{center}
\includegraphics[width=0.75\textwidth]{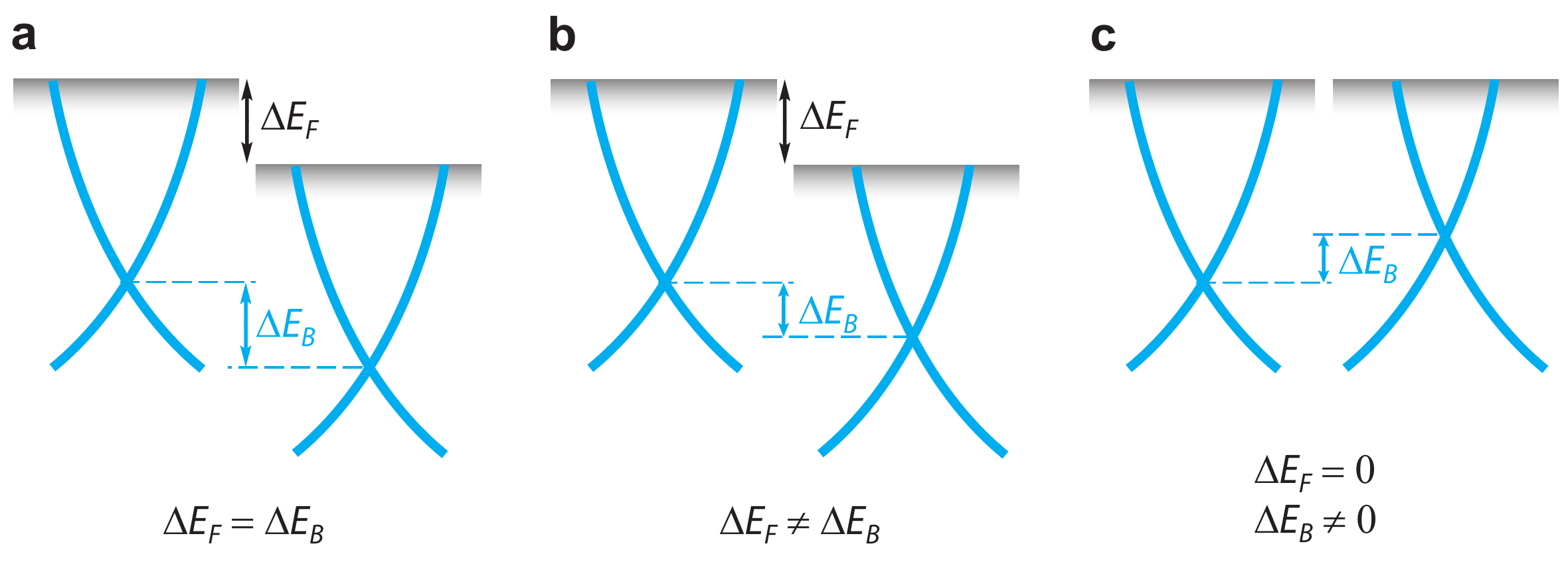}
\caption{
A classification of the photovoltage effects as seen by photoemission. (a) A so-called ``charging" of the whole sample due to absence of a good Ohmic contact between the surface of the sample and an electron analyser appears as a shift, $\Delta E_F$, of the Fermi level, $E_F$, of the sample under illumination in respect to its equilibrium position or to the $E_F$ of the analyser. (b) In the most general case the light induced photovoltage does both affect the surface charge region and create the charge of the sample. (c) In case of a highly conductive surface (and poorly conductive or insulating sample volume), its Fermi level remains equal to the one of the analyser and the only observed photovoltage effect is the surface states gating.
\label{fig2}}
\end{center}
\end{figure*}

Figure 2 presents the two key observations: (1) The number of the topological charge carriers depends on the excitation energy. (2) This dependence correlates with the total current of photo electrons. The $hv$-dependence of the binding energy of the Dirac point, or, in other words, of the surface state occupancy, can be seen on the raw ARPES spectra presented in panels (a) and (b) for a pure Bi$_2$Se$_3$ and a Cu- intercalated one, respectively. It is summarized in panel (c) for both samples in terms of the surface states Fermi surface area. While the effect is weak for the pure crystal, it appears as a well detectable step at about 30 eV for Cu$_{0.07}$Bi$_{2}$Se$_{3}$. The error bars include both the accuracy of the estimation of the Fermi surface area and its detected variations during the experiment.

Taking into account the 2D nature of the surface states, the observation (1) is very surprising. Despite the earlier reported time shift of the surface states that was associated with the band bending \cite{Hsieh}, absence of essential $hv$ dependence of those states has been considered as a proof of their surface origin \cite{Xia}. Unlike the exposition time dependent band bending, the described effect is much larger (150 meV vs 50 meV \cite{Hsieh}) and perfectly reproducible. The observed $hv$-dependence also cannot be explained by possible three-dimensionality of the bands because it is monotonic rather then oscillating. Actually, the earlier conclusion about surface state origin of the Dirac cone dispersion has been based on the measurements in the low excitation energy range, 19-31 eV. So, our result confirms the conclusion about two-dimensionality of the Dirac cone forming states on the basis of the data from the much wider energy range.

Ruling out the three-dimensionality issue, it is natural to assume that the observed change of the electronic occupation of the surface states is caused by the photoemission process. Due to discrete energy spectrum of electrons in atoms, the absorption of the light and, consequently, the photoemission current from the solids, exhibit a stair-like dependence on photon energy, with the steps, known as absorption edges, occurring at the binding energies of the core electrons plus the work function, $\phi$. Thus, one may expect an abrupt change in electron concentration at those energies. Indeed, the energy around 30 eV, at which the Dirac cone occupancy in the Cu-intercalated Bi$_2$Se$_3$ changes by a factor of two, corresponds to the lowest binding energy of the core electrons in this compound. The energy levels of those electrons are seen as two narrow peaks around 30 eV and can be unambiguously associated with the Bi 5d$_{5/2}$ and 5d$_{3/2}$ electrons residing at 23.8 and 26.9 eV binding energies \cite{Fuggle}, respectively ($\phi$ = 4.3 eV). The integral density of states (shown in Fig.~2c as a gray shaded area), that should be roughly proportional to the photocurrent, changes more than 10 times at this absorption step. This allows one to conclude that the observed change in the surface state occupancy is caused by a photovoltage effect.

The photovoltage effect on semiconducting surfaces and interfaces has been studied since the late 1940s \cite{Brattain} and the surface photovoltage method has been used as an extensive source of surface and bulk information on various semiconductors and semiconductor interfaces \cite{Garrett,Kronik} but, despite the great body of work, the microscopic description of the effect and related band bending is missing. In this regard, ARPES study of the topologically protected surface states can provide indispensible information for understanding the macroscopic parameters of the photovoltage effect starting from the electronic band structure of the surface states. On the other hand, the observed effect is peculiar since we see no detectable shift of the Fermi level of the sample with respect to the Fermi level of the spectrometer, which would be expected for a semiconducting sample with a grounded Ohmic back contact \cite{Kronik} (see Fig.~3). This can happen if some volume of the sample under the light spot is charged because of very low inter-layer conductivity while the surface of the crystal remains neutral due to good conductivity of the topmost layer, which may be a consequence of a topological protection of the surface states \cite{Chen} or a formation at the surface of a two-dimensional electron gas \cite{Bianchi}.

Naturally, the photo-induced gating should be material dependent and particularly sensitive to the inter-layer conductivity. One would expect that when the conductivity is lower the gating effect is higher. Thus, the intercalation, increasing the interlayer distance, decreases the conductivity and increases the gating effect. This is consistent with both the observed enhancement of the effect in Cu-intercalated Bi$_2$Se$_3$ as compared to the pure Bi$_2$Se$_3$ and with the resistivity values measured for these crystals. We note, however, that it is difficult to make quantitative comparison between ARPES and transport since the transport measurements provide us the bulk information while the effect appears in a very narrow surface region in which the local conductivity should be different due to band bending. Also, it is clear that, due to a number of parameter involved, the microscopic understanding of this effect requires thorough investigation of its dependence on photon flux, temperature, and doping. In this communication, we have intended to show the robustness of the gating effect in Cu$_{0.07}$Bi$_{2}$Se$_{3}$ and its correlation with the photo-induced current. Also we note that the accuracy of present experiment, see error bars in Fig.~2c, does not allow us to conclude about the value of the effect for the pure crystals. In this case, it is important to add that a small photovoltage effect for the epitaxial Bi$_2$Se$_3$ films have been very recently discussed in Ref. \onlinecite{Zhang}.

In summary, we observe the effect of photoemission induced gating of the topological surface states on Cu$_x$Bi$_{2}$Se$_{3}$ that may stimulate the use of the topological insulators in electronics \cite{Nayak,Zutic}. The observed enhancement of the effect by Cu intercalation shows the way to control it from the material side. While the peculiarities caused by the presence of the topologically protected surface states have to be understood, the very fact that the photovoltage effect has been observed directly for the compound in which the surface states dispersion can be measured in details and controlled opens opportunity to study the microscopic mechanisms of the surface photovoltage effects on semiconducting surfaces and interfaces.

We acknowledge discussions with Eugene Krasovskii, Alexander Yaresko, Vladislav Kataev, and Alexander Lavrov. The project was supported by the DFG under Grants No. KN393/4, BO 1912/2-1, and priority programme SPP 1458.

%Correspondence and requests for materials should be addressed to A.A.K. (a.kordyuk@ifw-dresden.de).

\onecolumngrid

\begin{thebibliography}{10}

\bibitem{Xia} Y. Xia, D. Qian, D. Hsieh, L. Wray, A. Pal, H. Lin, A. Bansil, D. Grauer, Y. S. Hor, R. J. Cava, and M. Z. Hasan, Nature Phys. \textbf{5}, 398-402 (2009)
\bibitem{Hsieh} D. Hsieh, Y. Xia, D. Qian, L. Wray, J. H. Dil, F. Meier, J. Osterwalder, L. Patthey, J. G. Checkelsky, N. P. Ong, A. V. Fedorov, H. Lin, A. Bansil, D. Grauer, Y. S. Hor, R. J. Cava, and M. Z. Hasan, Nature \textbf{460}, 1101-1105 (2009).
\bibitem{Chen} Y. L. Chen, J. G. Analytis, J.-H. Chu, Z. K. Liu, S.-K. Mo, X. L. Qi, H. J. Zhang, D. H. Lu, X. Dai, Z. Fang, S. C. Zhang, I. R. Fisher, Z. Hussain, and Z.-X. Shen, Science \textbf{325}, 178-181 (2009).
\bibitem{Moore} J. E. Moore, Nature \textbf{464}, 194-198 (2010).
\bibitem{Kane} C. L. Kane, Nature Phys. \textbf{4}, 348-349 (2008).
\bibitem{Fu} L. Fu, C. L. Kane, and E. J. Mele, Phys. Rev. Lett. \textbf{98}, 106803 (2007).
\bibitem{Wilczek} F. Wilczek, Nature Phys. \textbf{5}, 614-618 (2009).
\bibitem{Nayak} C. Nayak, S. H. Simon, A. Stern, M. Freedman, and S. D. Sarma, Rev. Mod. Phys. \textbf{80}, 1083-1159 (2008).
\bibitem{Zutic} I. \v{Z}uti\'{c}, J. Fabian, and S. Das Sarma, Rev. Mod. Phys. \textbf{76}, 323-410 (2004).
%\bibitem{Loudon} R. Loudon, Quantum Theory of Light 3rd edn, Ch. 1 (Oxford Univ. Press, 2000).
\bibitem{Hor} Y. S. Hor, A. J. Williams, J. G. Checkelsky, P. Roushan, J. Seo, Q. Xu, H. W. Zandbergen, A. Yazdani, N. P. Ong, and R. J. Cava, Phys. Rev. Lett. \textbf{104}, 057001 (2010).
\bibitem{Wray} L. Wray, S. Xu, J. Xiong, Y. Xia, D. Qian, H. Lin, A. Bansil, Y. Hor, and R. J. Cava, arXiv:0912.3341v1 (2009).
\bibitem{Geim} A. K. Geim and K. S. Novoselov, Nature Mater. \textbf{6}, 183-191 (2007).
\bibitem{Steinberg} H. Steinberg, D. R. Gardner, Y. S. Lee, and P. Jarillo-Herrero, arXiv:1003.3137 (2010).
\bibitem{Hyde} G. R. Hyde, H. A. Beale, I. L. Spain, and J. A. Woollam, J. Phys. Chem. Solids \textbf{35}, 1719 (1974).
\bibitem{Vasko} A. Va\v{s}ko, L. Tich\'{y}, J. Hor\'{a}k, and J. Weissenstein, Appl. Phys. \textbf{5}, 217-221 (1974).
\bibitem{Borisenko} S. V. Borisenko, V. B. Zabolotnyy, D. V. Evtushinsky, T. K. Kim, I. V. Morozov, A. N. Yaresko, A. A. Kordyuk, G. Behr, A. Vasiliev, R. Follath, and B. Buechner, Phys. Rev. Lett. \textbf{105}, 067002 (2010).
\bibitem{Fuggle} J. C. Fuggle and N. M{\aa}rtensson, J. Electron Spectrosc. Relat. Phenom. \textbf{21}, 275-281 (1980).
\bibitem{Brattain} W. H. Brattain, Phys. Rev. \textbf{72}, 345-345 (1947).
\bibitem{Garrett} C. G. B. Garrett and W. H. Brattain, Phys. Rev. \textbf{99}, 376-387 (1955).
\bibitem{Kronik} L. Kronik and Y. Shapira, Surf. Sci. Rep. \textbf{37}, 1-206 (1999).
\bibitem{Bianchi} M. Bianchi, D. Guan, S. Bao, J. Mi, B. B. Iversen, P. D. C. King, and P. Hofmann, Nature Commun. \textbf{1}, 128 (2010).
\bibitem{Zhang} Y. Zhang, K. He, C.-Z. Chang, C.-L. Song, L.-L. Wang, X. Chen, J.-F. Jia, Z. Fang, X. Dai, W.-Y. Shan, S.-Q. Shen, Q. Niu, X.-L. Qi, S.-C. Zhang, X.-C. Ma, and Q.-K. Xue, Nature Phys. \textbf{6}, 584 (2010).

\end{thebibliography}
\end{document}